\documentclass{emulateapj}
\newcommand{\cote}{C\^{o}t\'{e}\ }
\newcommand{\jordan}{Jord\'{a}n\ }
\newcommand{\hasegan}{Ha{\c s}egan\ }
\newcommand{\etal}{et~al.\ }
\slugcomment{Accepted for publication in The Astrophysical Journal}

\shorttitle{Alignment of GC Systems}
\shortauthors{Wang et al.}

\begin{document}

%% LaTeX will automatically break titles if they run longer than
%% one line. However, you may use \\ to force a line break if
%% you desire.

\title{The ACS Virgo Cluster Survey. XVII. The Spatial Alignment of Globular
  Cluster Systems with Early-Type Host Galaxies\altaffilmark{*}}

%% Use \author, \affil, and the \and command to format
%% author and affiliation information.
%% Note that \email has replaced the old \authoremail command
%% from AASTeX v4.0. You can use \email to mark an email address
%% anywhere in the paper, not just in the front matter.
%% As in the title, use \\ to force line breaks.

\author{Qiushi Wang\altaffilmark{1,2}}

\author{Eric W. Peng\altaffilmark{1,3,4}}

\author{John P. Blakeslee\altaffilmark{5}}

\author{Patrick C\^{o}t\'{e}\altaffilmark{5}}

\author{Laura Ferrarese\altaffilmark{5}}

\author{Andr\'{e}s Jord\'{a}n\altaffilmark{6}}

\author{Simona Mei\altaffilmark{7,8}}

\author{Michael J. West\altaffilmark{9,10}}

%% Notice that each of these authors has alternate affiliations, which
%% are identified by the \altaffilmark after each name.  Specify alternate
%% affiliation information with \altaffiltext, with one command per each
%% affiliation.

\altaffiltext{*}{Based on observations with the NASA/ESA {\it Hubble
    Space Telescope} obtained at the Space Telescope Science Institute,
  which is operated by the Association of Universities for Research in
  Astronomy, Inc., under NASA contract NAS 5-26555.}
\altaffiltext{1}{Department of Astronomy, Peking University, Beijing
  100871, China}
\altaffiltext{2}{Physics Department, New York University, 4 Washington
  Place, New York, NY 10003, USA}
\altaffiltext{3}{Kavli Institute for Astronomy and Astrophysics,
  Peking University, Beijing 100871, China}
\altaffiltext{4}{Corresponding author; peng@pku.edu.cn}
\altaffiltext{5}{Herzberg Institute of Astrophysics,
  National Research Council of Canada,
  5071 West Saanich Road, Victoria, BC  V9E 2E7, Canada}
\altaffiltext{6}{Departamento de Astronom\'{\i}a y Astrof\'{\i}sica,
Pontificia Universidad Cat\'olica de Chile, Casilla 306, Santiago 22,
Chile}
\altaffiltext{7}{University of Paris 7 Denis Diderot,  75205 Paris Cedex
  13, France}
\altaffiltext{8}{GEPI, Observatoire de Paris, Section de Meudon, 5 Place
  J. Janssen, 92195 Meudon Cedex, France}
\altaffiltext{9}{Maria Mitchell Observatory,
  4 Vestal Street, Nantucket, MA 02554, USA}
\altaffiltext{10}{European Southern Observatory, Alonso de Cordova 3107,
 Vitacura, Santiago, Chile}

%% Mark off your abstract in the ``abstract'' environment. In the manuscript
%% style, abstract will output a Received/Accepted line after the
%% title and affiliation information. No date will appear since the author
%% does not have this information. The dates will be filled in by the
%% editorial office after submission.

\begin{abstract}
We study the azimuthal distribution of globular clusters (GCs)
in early-type galaxies and compare them to their host galaxies using
data from the ACS Virgo Cluster Survey. We find that in host galaxies
with visible elongation ($\epsilon > 0.2$) and intermediate to high
luminosities ($M_z<-19$), the GCs are preferentially aligned along the
major axis of the stellar light.  The red (metal-rich) GC
subpopulations show strong
alignment with the major axis of the host galaxy, which supports the
notion that these GCs are associated with metal-rich field stars.
The metal-rich GCs in lenticular galaxies show signs of being more
strongly associated with disks rather than bulges.
Surprisingly, we find that the blue (metal-poor) GCs can also show the
same correlation.  If the metal-poor GCs are part of the early formation of the
halo and built up through mergers, then our results support a picture
where halo formation and merging occur anisotropically, and where the
present day major axis is an indicator of the preferred merging axis.
\end{abstract}

%% Keywords should appear after the \end{abstract} command. The uncommented
%% example has been keyed in ApJ style. See the instructions to authors
%% for the journal to which you are submitting your paper to determine
%% what keyword punctuation is appropriate.

%% Authors who wish to have the most important objects in their paper
%% linked in the electronic edition to a data center may do so in the
%% subject header.  Objects should be in the appropriate "individual"
%% headers (e.g. quasars: individual, stars: individual, etc.) with the
%% additional provision that the total number of headers, including each
%% individual object, not exceed six.  The \objectname{} macro, and its
%% alias \object{}, is used to mark each object.  The macro takes the object
%% name as its primary argument.  This name will appear in the paper
%% and serve as the link's anchor in the electronic edition if the name
%% is recognized by the data centers.  The macro also takes an optional
%% argument in parentheses in cases where the data center identification
%% differs from what is to be printed in the paper.

%% From the front matter, we move on to the body of the paper.
%% In the first two sections, notice the use of the natbib \citep
%% and \citet commands to identify citations.  The citations are
%% tied to the reference list via symbolic KEYs. The KEY corresponds
%% to the KEY in the \bibitem in the reference list below. We have
%% chosen the first three characters of the first author's name plus
%% the last two numeral of the year of publication as our KEY for
%% each reference.

\section{Introduction}
\setcounter{footnote}{0}

The classical illustration of a globular cluster system is as a
spherical halo population around the host galaxy.  Galaxies
themselves, however, can have a wide variety of shapes, from
nearly-spherical to highly anisotropic.  There are many reasons why GC
systems and stellar halos could be spherical.  They could have formed
very early in a chaotic fashion, similar to the classic ``monolithic
collapse'', before dissipation could collect gas
into a disk.  They could also have originally formed in disks but then had
their orbits distributed into a spherical halo via major mergers
(e.g., Toomre \& Toomre 1972).
It is also possible that at least some halo stars and GCs have been
accreted from a more spherically distributed population of dwarf-like
galaxies and halos (e.g., Searle \& Zinn 1978; \cote, Marzke, \& West
1998). 

It is not necessary, however, that in all cases the shape of the
stellar halo and its GC system be decoupled from its
host.  Mergers along a preferred axis in a galaxy cluster can produce
alignment between the major axis of the brightest cluster galaxy (BCG)
and the cluster major axis, as can intrinsic elongation of the dark
matter halo (Binggeli 1982; West 1994; Hashimoto, Henry \& Boehringer
2008).  The distribution of satellites is also not necessarily isotropic around galaxies, as
alignments between satellites and host galaxies have been found in
many studies, although not always in the same sense.  Holmberg (1969)
and others (e.g., Zaritsky \etal 1997)
found that satellites of disk galaxies tend to be close to the minor axes
of their hosts, but other studies (e.g., Brainerd 2005; Yang \etal
2006; Bailin \etal 2008) have found that satellites are preferentially
along the major axes of galaxies, particularly early-types.

The oldest stars in the Universe are found in stellar halos and GCs,
and these populations are the oldest visible collisionless tracers
within galaxies.  The shapes of 
halo stellar populations tells us about the merging and accretion
history of the galaxy over a Hubble time.
Studying the outer shapes of stellar halos, however, is extremely difficult because of
their low surface brightnesses.  Globular clusters, however, are
readily identified out to large distances using the Hubble Space
Telescope ($D\lesssim100$~Mpc, e.g., Peng \etal 2011), and are
present in nearly every galaxy except the faintest dwarf galaxies.
GC systems in massive galaxies are also known to have bimodal color
distributions, which may correspond to ``metallicity
subpopulations'' whose mean metallicities correlates with the mass of
the host galaxy (e.g., Larsen \etal 2001; Peng \etal 2006a).  Although
the exact interpretation of these color distributions in metallicity
is currently debated (c.f., Yoon \etal 2006, 2011), it is still true that
with reasonably deep imaging, GCs can be easily detected, and can
provide some chemical information on the underlying stellar population.

%%%%%%%%%%%%%%%%%%%%%%%%%%%%%%%%%%%%%%%%%%%%%%%%%%%%%%%%%%%%%%%%%%%%%%%%%%
\begin{figure*}[ht]
%\centering
%\begin{picture}(240,115)(0,0)
%\put(0,0){\includegraphics[width=4cm,height=4cm]{imgs/mine/galaxy/1692all}}
%\put(130,0){\includegraphics[width=4cm,height=4cm]{imgs/mine/galaxy/1692allzoom}}
%\end{picture}
\plottwo{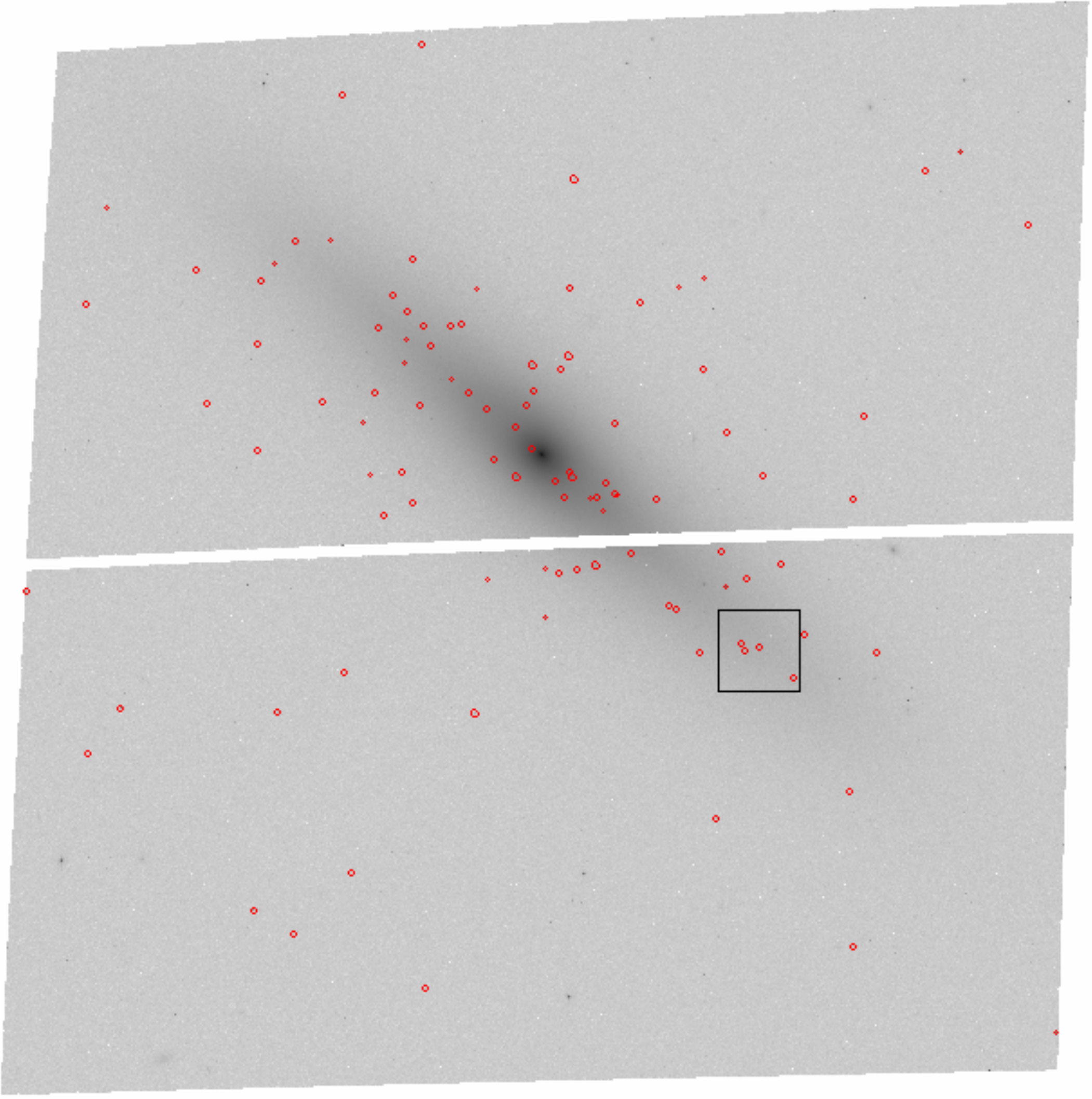}{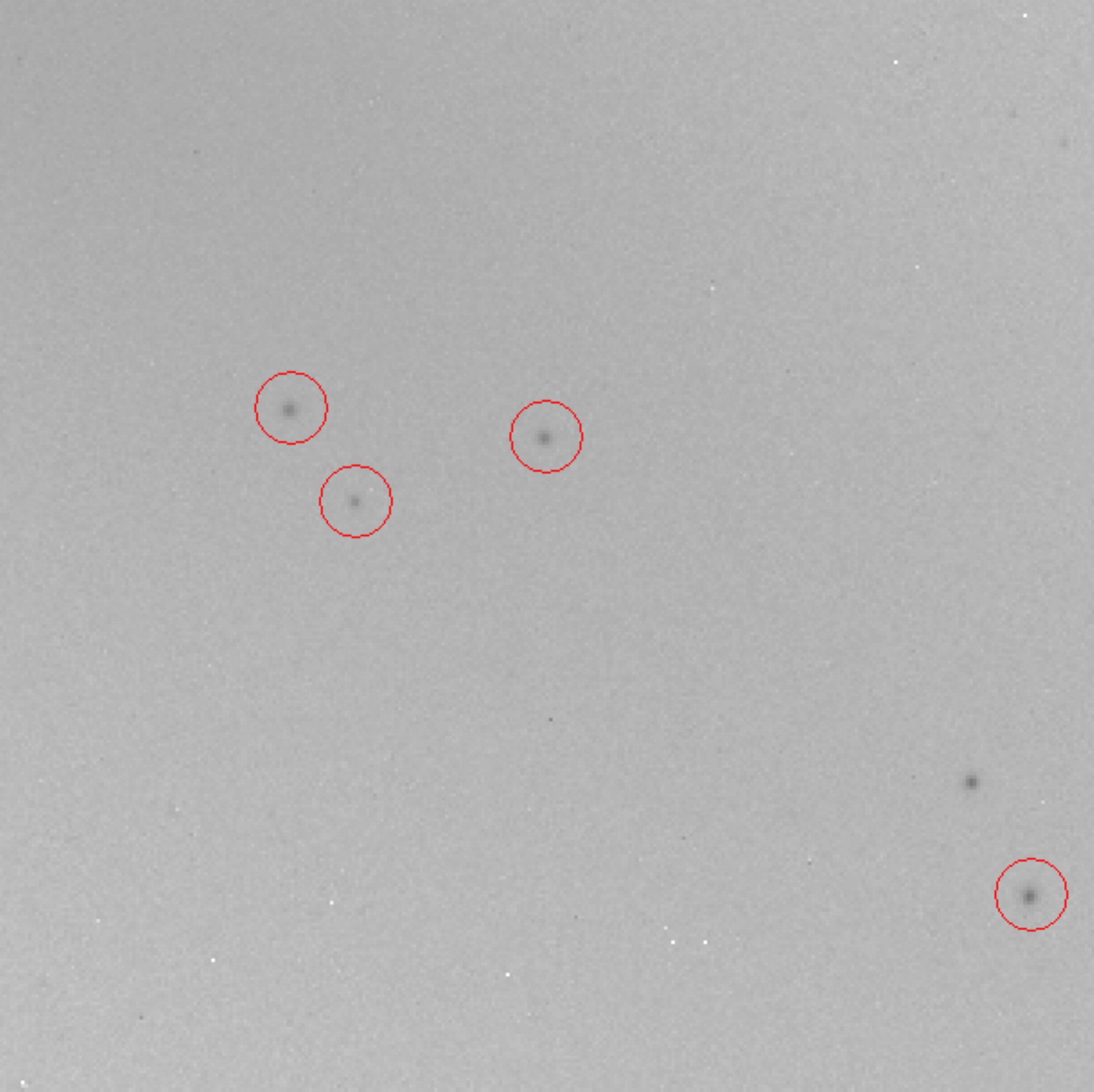}

\caption{Left panel: VCC~1692 and globular cluster candidates with $class>0.9$.
The white gap in the middle of the image is the gap between
two WFC detectors. Right panel: enlarged image of the squared aera in the left.\label{fig:1692zoom}}
\end{figure*}
%%%%%%%%%%%%%%%%%%%%%%%%%%%%%%%%%%%%%%%%%%%%%%%%%%%%%%%%%%%%%%%%%%%%%%%%%%

Recent studies of extragalactic GC systems have taken advantage of
surveys of early-type galaxies using {\it HST}, particularly in the
Virgo and Fornax galaxy clusters.  In this paper, we undertake a
study of the azimuthal distributions of GC systems in early-type
galaxies using data from the ACS Virgo Cluster Survey (\cote
\etal 2004).  The high spatial resolution and depth of this survey has
already been used to study GC color distributions (Peng \etal 2006a),
size distributions (\jordan \etal 2005), luminosity functions (\jordan
\etal 2006,2007), formation efficiencies (Peng \etal 2008), and
color-magnitude relations (Mieske \etal 2006, 2010), as well as other dense
stellar systems such as diffuse star clusters
(Peng \etal 2006b), ultra-compact dwarfs (\hasegan \etal 2005) and
nuclear star clusters (\cote \etal 2006; \cote \etal 2007). This data set is
well-suited to the study of star clusters in Virgo early-type
galaxies, and we use the GC catalog of \jordan \etal (2009).

In this paper, we seek to answer the
questions: 1) Can GC systems be non-spherical?  2) If so, does the
major axis of the GC system have any relationship with that of the
host galaxy?  3) Do the blue and red GC subpopulations have different
distributions around their hosts?  The GC systems of a few massive
galaxies have been shown to be flattened (e.g.\ McLaughlin \etal 1994, 1995),
but in this paper, we present the first large study of GC systems and their
possible alignment with their host galaxies.

\section{Data and Methods}

\subsection{Globular Cluster Catalog}
The ACS Virgo Cluster Survey (ACSVCS; \cote \etal
2004) observed one hundred Virgo Cluster early-type galaxies in
two filters F475W ($g$) and F850LP ($z$). We use the globular cluster
catalog for all ACSVCS galaxies presented by \jordan \etal (2009),
which gives their positions, photometry, half-light radii, and a
{\it class} parameter which indicates the probability a given object
is a GC (see \jordan \etal 2004 for the specifics of the data reduction).
By the virtue of the high resolution of Hubble Space
Telescope (HST), we can measure the total magnitude $z_0$ and half light
radius $r_\mathrm{h}$ of every globular cluster candidate. The GC
probability is determined using these parameters in comparison to
control fields customized for the observational depth for each galaxy
(see Peng \etal 2006a, Figure~1).  In this paper, we select objects
with $class > 0.9$ as GCs.  This is more similar to the stringent selection used
in the GC system color gradient study of Liu \etal (2011), which also
required strict selection against background contaminants. For the
dwarf galaxies in the sample (where contamination is highest and
potentially a larger issue), we estimate from control fields that
there should be only $\sim1.5$ contaminants per galaxy selected by
our size-magnitude-color criteria. This extremely low level of
contamination minimizes the effect of anisotropy in the background.

This selection does not include the
objects discussed as ``faint fuzzies'' in Larsen \& Brodie (2000) or
``diffuse star clusters'' (DSCs) in Peng \etal (2006b).  In cases
where a distance is needed, we use the SBF distances presented in
Blakeslee (2009, also see Mei \etal 2007).  We use a mean distance of
16.5~Mpc to the Virgo Cluster (Mei \etal 2005).

Several galaxies in the ACSVCS sample are in close proximity to large
neighboring galaxies, and their GC systems are either overwhelmed by
the GC system of their neighbor, or strongly depleted in GCs, possibly due to tidal stripping.
We exclude these six galaxies from our analysis: VCC~1327, VCC~1297, VCC~1279,
VCC~1938, VCC~1192 and VCC~1199.

\subsection{Methodology}

The relatively wide field of view of the ACS/WFC ($202\arcsec\times
202\arcsec$, or $16.2\times16.2$~kpc), allows us to study the
azimuthal distribution of GC systems in all but a few of the more massive
galaxies in the ACSVCS sample. The image of one ACSVCS galaxy, VCC~1692, is shown in
Figure~\ref{fig:1692zoom}.  This lenticular galaxy is the twelfth brightest
galaxy in the ACSVCS sample, and provides a nice
example to illustrate our analysis methods.  As can be seen even by eye, the
globular clusters in VCC~1692 tend to align with the galaxy light, and
are preferentially clustered close to the plane of the galaxy
(Figure~\ref{fig:1692zoom}, left).  Our goal is to quantify this, particularly in
galaxies where the geometry may not be so favorable.

To do this, we analyze the distribution of GCs in azimuth (position
angle, $\phi$).  This approach, rather than a more complicated fitting
of isopleths, is applicable to a wide range of data, from the relative
large GC system shown in Figure~\ref{fig:1692zoom} to the smaller ones
belonging to dwarf galaxies.  We use the azimuthal distribution to test
for departures from circular symmetry.  In cases where the azimuthal
distribution appears anisotropic, we can also test for alignment
between the major axis of the GC system and that of the galaxy.

This test, looking for departures from uniformity in the projected
azimuthal distribution, is in many ways the simplest one we can
do. Understanding the intrinsic shapes of GC system is a much more
difficult question that would require incorporating inclination, not
to mention many more intrinsic parameters. This paper is a simple, but
interesting first step to generally study anisotropic spatial
distributions in GC systems. 

We analyze the azimuthal distributions of GCs using two methods.
First, we use the Kolmogorov-Smirnov (K-S) test to test for departures
from a random (isotropic) distribution.  Second, we use binned
distributions in azimuth to see if the GC system has a preferred axis.

\subsection{Generating Random Comparison Samples}

To perform these tests, we need to create control samples that mimic a
random azimuthal distribution of GCs (the equivalent to an
isotropic, or spherical, GC system).
Because the boundary of the image is not a circle, and the center
of the galaxy is not exactly at the position of the image center, the
observed area in a given $\Delta\phi$ changes with azimuth
($\phi$).  Therefore, an azimuthally random distribution of GCs
about the galaxy will not result in a constant observed number per unit
azimuth.  Taking a mean surface number density as a function of $\phi$ is
also problematic because this density changes as function of projected
radius from the galaxy center.

For each galaxy, we create a ``randomized'' sample of GCs purely in
azimuth, bypassing the need to know the radial density profile of the
GC system.  For each GC, we first assign it a random angle position while
keeping its projected radius unchanged. We then see if the randomized
position is within the image area of the galaxy. If not, we randomize
it again as in the first step, until the resulting globular cluster is
located within the image.  We continue to do this for each GC in the
sample until all GCs have new positions within the observed area of
the image, creating a fully randomized control sample.  By
``randomized'', we mean that the GCs in the image randomly populate
the  distribution in aziumuth, modulo the varying image area,
and without needing to make assumptions about the GC 
radial density profile. We define $N$ 
as the number of control samples we generate. Unless mentioned
otherwise, $N$ in our study equals one thousand.

Some of the objects classified as GCs will in fact be contamination
from background galaxies.  Assuming that background objects are
distributed isotropically in azimuth, their only effect is to dilute
the signal of any alignment within the GC system.  Including them in
the randomized samples should not introduce a bias unless they are not
isotropically distributed in the field of view.  Fortunately,
the GC samples are already quite clean of contaminants, with an
expectation of only $\sim1.5$ background objects per galaxy (Liu \etal
2011). 

We also analyze the azimuthal distributions of the GC color
subpopulations.  We divide the GCs into blue and red populations
using their $(g-z)$ color.  For the more massive GC systems, we use
the two-Gaussian mixture model fits from Peng \etal (2006a) to
determine the crossover color between the two populations.  For
galaxies where there are fewer GCs, we choose a color of
$(g-z)=1.10$~mag to separate blue from red.  We ultimately performed all
of our analysis using a stricter separation between the two
populations ($(g-z)<1.00$ for blue GCs and $(g-z)>1.20$ for red GCs.
None of our conclusions change when we change the color cuts.  When
analyzing the color subpopulations, we generate separate randomized
samples for each one.

%%%%%%%%%%%%%%%%%%%%%%%%%%%%%%%%%%%%%%%%%%%%%%%%%%%%%%%%%%%%%%%%%%%%%%%%%%%%%
\begin{figure}[t]
\plotone{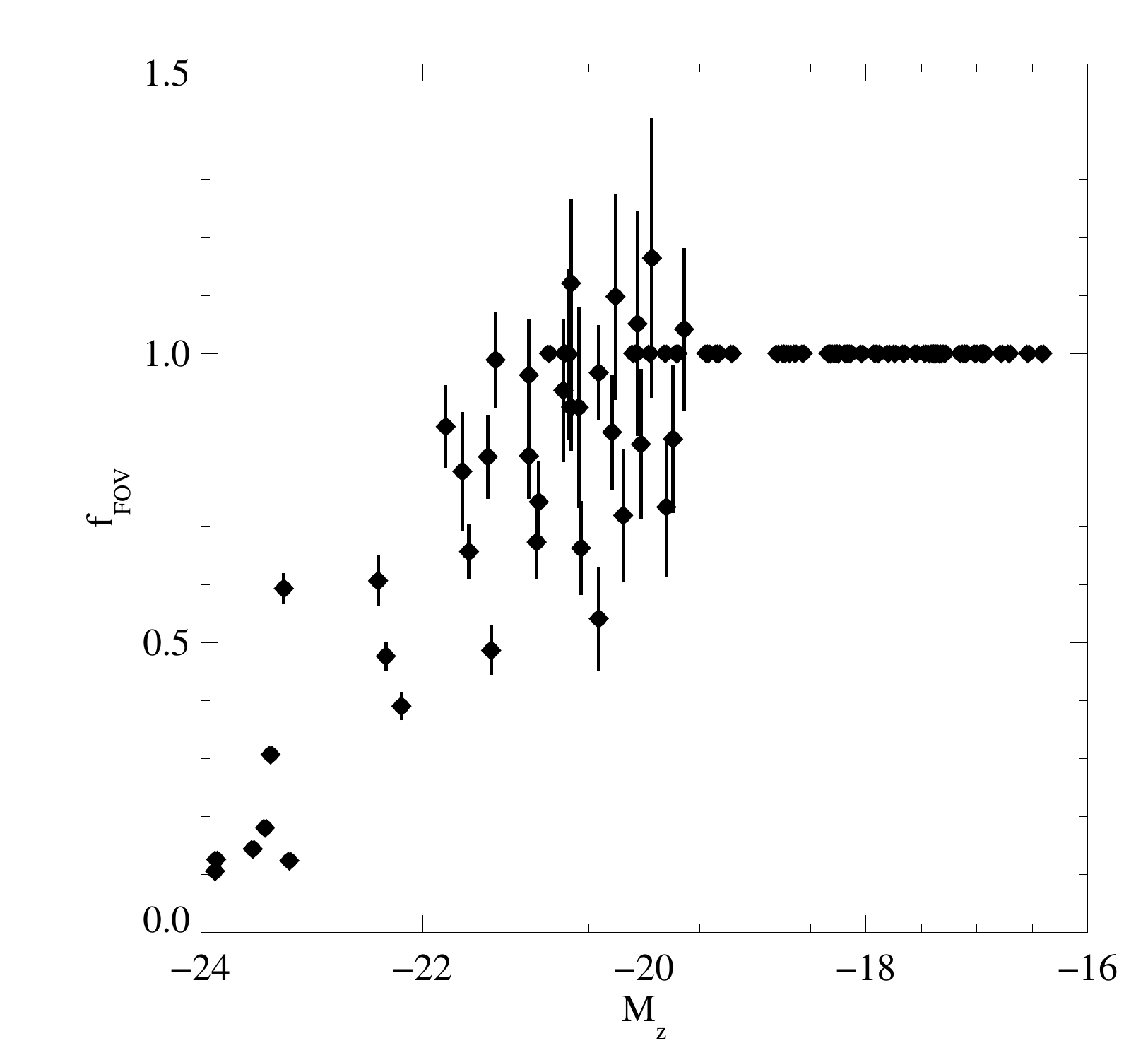}

\caption{For the more luminous galaxies in our sample, the ACS/WFC
  field of view covers only a fraction of the full GC system. We plot
  this fraction against $M_z$ to show that for systems with
  $M_z\lesssim -20$~mag, we are missing a measureable fraction of
  GCs. Galaxies for which the ratio is exactly unity are those for
  which ACS/WFC field is presumed to cover the entire GC
  system. Values larger than unity are due to errors in fitting the
  total number of GCs from the GC radial density profile. Only 9 of
  the 94 galaxies in the sample have less than 50\% coverage.
\label{fig:fov}}
\end{figure}
%%%%%%%%%%%%%%%%%%%%%%%%%%%%%%%%%%%%%%%%%%%%%%%%%%%%%%%%%%%%%%%%%%%%%%%%%%%%%

\subsection{The HST/ACS Field of View}

One of the main limitations of our study is the ACS/WFC field of
view. This characteristic of the ACSVCS sample has been described in
previous studies (particularly in Peng et al.\ 2008, Section~3), but
we further quantify the issue in this paper. For most of the
early-type dwarf galaxies in the sample, the
$202\arcsec\times202\arcsec$ ACS field of view is more than adequate
to cover the entire GC system, but this is not the case for
galaxies with $M_z\lesssim-20$~mag. We illustrate this in
Figure~\ref{fig:fov}, where we show the fraction of the entire GC
system included in the ACS/WFC field of view. This total number of GCs
for larger GC systems was determined using a S\'ersi\'c fit to the
spatial density profile (Peng et al.\ 2008). Where the ratio is
exactly unity (for most dwarfs), the total number was determined to be the number
observed in the ACS/WFC field. For the most luminous galaxies in our
sample, only $\sim10\%$ of the GCs are being analyzed. Nevertheless,
the quality of the data is high and we can still draw interesting
conclusions from these inner regions. Most of our sample, however, is
not particularly compromised; in only nine of our 94 galaxies do we
sample less than half of the GC system. Similar analyses of the full GCS
at large radii will require wide-field imaging, like those in the Next
Generation Virgo Survey (Ferrarese \etal 2012).

\section{Results}
\subsection{The Kolmogorov-Smirnov Test}

\begin{figure}[t]
\plotone{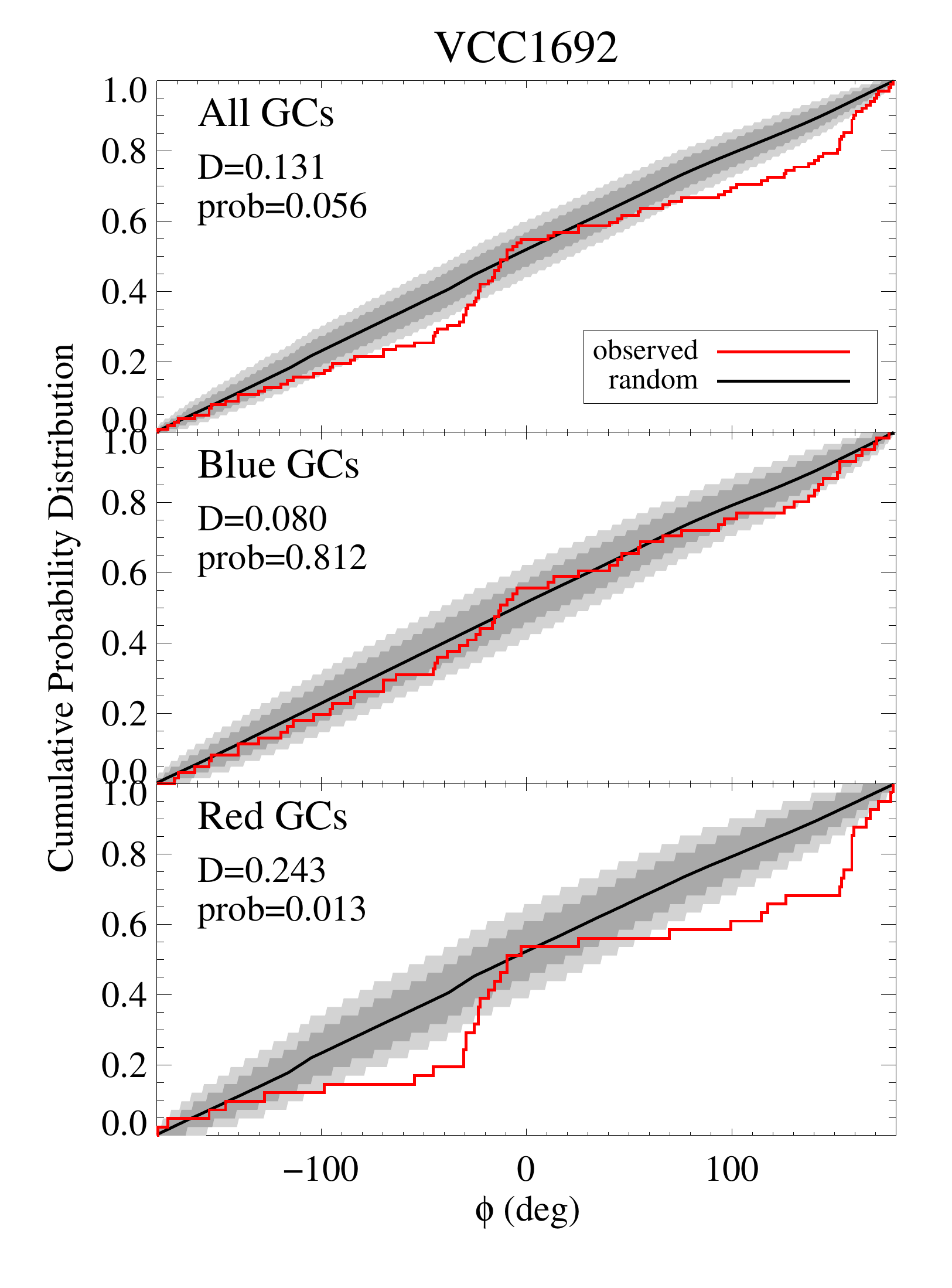}

\caption{VCC1692. The top, middle and bottom plots show the
cumulative probability distributions (CPDs) of GCs versus position
angle (degrees east from north) for all, blue and red GCs,
respectively. The red lines represent the observed distribution, and
the black lines represent the sum of 10000 randomized samples. The dark
gray and light gray regions are the regions encompassing
68\% and 90\%, respectively, of all randomly generated samples. The
values for the $D$ statistic and the probability that the two
samples were drawn from the same distribution are labeled in each
plot.  Note the large excesses above the random distribution at
$-35^\circ$ and $+145^\circ$.  These correspond to GCs
preferentially clustered around the galaxy major axis.  The effect
is more pronounced for the red GCs.  \label{1692}}
\end{figure}

We employ the two-sided Kolmogorov-Smirnov test to quantitatively study the
alignment between GC systems and their host galaxies. Specifically,
for each galaxy, first we create randomized samples as described above,
then we use the Kolmogorov-Smirnov test to compare the
randomized distribution with the observed one. The result can be
displayed visually in the cumulative distributions of both observed
and randomized samples. These distributions for the galaxy VCC1692 are
shown in figure \ref{1692}. There are two evident knee-like steps in
the red GCs' cumulative azimuthal distribution, 
spaced $180^\circ$ apart, which indicates an alignment along an
axis. The cumulative distribution of the blue GCs also show these two
steps, but to a lesser degree.

%%%%%%%%%%%%%%%%%%%%%%%%%%%%%%%%%%%%%%%%%%%%%%%%%%%%%%%%%%%%%%%%%%%%%%%%%%%%%
%%%%%%%%%%%%%%%%%%%%%%%%%%%%%%%%%%%%%%%%%%%%%%%%%%%%%%%%%%%%%%%%%%%%%%%%%%%%%

The two-sided K-S test generates the $D$ statistic, which is a measure
of the maximum deviation of one cumulative probability distribution from
another.  This statistic results in a probability that the two
data sets are drawn from the same distribution.  We have calculated
the distribution of $D$ statistic values using our observed data and
each of the 10000 azimuthally randomized samples. We find 
that this distribution is well-behaved and nearly Gaussian. In the case of
VCC~1692, there is only a $5.6\%$ probability that its GCs are
drawn from a random azimuthal distribution.  The probability is even
lower, $1.3\%$ for the red GCs alone.  The distribution of blue GCs is
much closer to that of the randomized sample, with a $81.2\%$ of
being drawn from a random distribution.

%%%%%%%%%%%%%%%%%%%%%%%%%%%%%%%%%%%%%%%%%%%%%%%%%%%%%%%%%%%%%%%%%%%%%%%%%%%%%%
\begin{figure}

\plotone{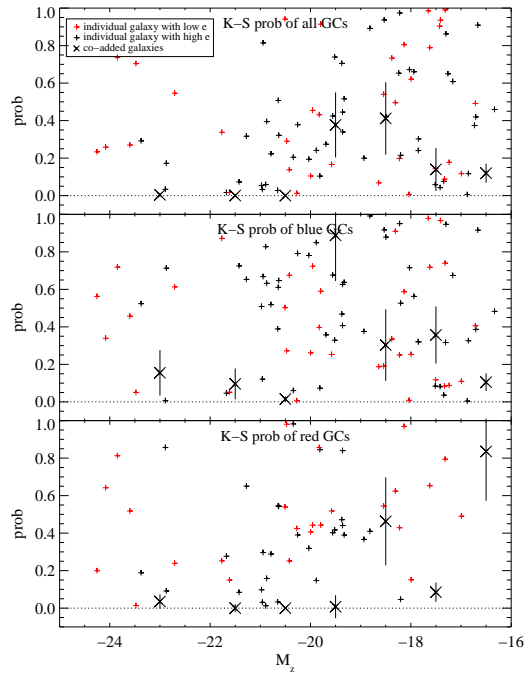}
\caption{K-S test results for individual galaxies and binned galaxy
  samples.  These figures plot the probability that the azimuthal
  distribution of GCs is drawn from a random distribution ($prob$)
  against the total luminosity of the galaxy.  Results are plotted
  for all GCs (top), blue GCs (middle), and red GCs (bottom).  Small
  red points are galaxies with ellipticities, $e$, less than
  0.2.  Small black points are galaxies with $e>0.2$.  Large
  points with error bars represent binned samples of galaxies with
  $e>0.2$.  For all samples of GCs, the more luminous galaxies,
  on average, have anisotropic GC systems.\label{fig:ksall}}
\end{figure}
%%%%%%%%%%%%%%%%%%%%%%%%%%%%%%%%%%%%%%%%%%%%%%%%%%%%%%%%%%%%%%%%%%%%%%%%%%%%%%

In Figure~\ref{fig:ksall}, we plot the results of the K-S test for all
galaxies in the sample. For each galaxy, we plot $prob$,
$prob_{blue}$, and $prob_{red}$, which are the probabilities that the entire
GC system, the blue GCs, and the red GCs are drawn from a random
distribution, respectively.  We plot these probabilities as a function
of the absolute $z$ magnitude of the galaxy ($M_z$). The galaxy
magnitudes are from Ferrarese \etal (2006) and Peng \etal (2008).

Figure~\ref{fig:ksall} shows that there is a large scatter in
probabilities for the ACSVCS galaxies.  The brightest galaxies suffer
from a small field of view, and the less luminous galaxies have few
GCs, creating increased scatter.  For some galaxies, however,
particularly those with $-23<z<-21$~mag, the probability that the GCs are
distributed in random azimuth is very small.  These galaxies tend to
be lenticular or elliptical galaxies with high axis ratio.  To
highlight this, we plot with different colors the galaxies with
ellipticities $\epsilon >0.2$, which are taken from the isophotal fitting of
Ferrarese \etal (2006, see their Figure~121 for the relationship between
ellipticity and luminosity).

Since some galaxies are so faint that there are too few detected
globular clusters to determine the trends, we bin galaxies by their
magnitudes and run the Kolmogorov-Smirnov test on the collective
data sets. If GC systems are possibly aligned with the shapes of their
host galaxies, then non-spherical galaxies provide a stronger test of
this hypothesis, so we only include galaxies with $\epsilon >0.2$ in the
binned samples.
When combining galaxies, we always use a GC's position angle from the
galaxy major axis.  This should enhance the signal if GC systems are
aligned with the galaxy light, but would decrease the signal if GC
systems are randomly oriented with respect to their hosts.  We use the
major axis position angles given by Ferrarese \etal (2006), averaging the values given for the $g$ and $z$ bands.

Furthermore, we use the bootstrap to determine the errors
in the probabilities produced by the Kolmogorov-Smirnov test. The
results are shown in Table~\ref{kst}.  In Figure~\ref{fig:ksall}, we
plot the result of
Kolmogorov-Smirnov test for individual galaxies and binned samples
(with error bars). For the binned samples of galaxies with $\epsilon >0.2$ in
particular, we can clearly see the trends of GC systems.  The most
obviously anisotropic distributions of GCs are for the red GCs in
the four most luminous galaxy bins.  For this magnitude range
($M_z\lesssim -19$~mag), on average, the probability that the GCs are isotropically
distributed in azimuth about the galaxy is near zero.  This is also
true, however, for the blue GCs in the three most luminous bins
($z\lesssim -20$~mag).  Once we get down to dwarf galaxy luminosities, the
alignment between the GC systems and the galaxies is less evident.
This might be because dwarfs have more spherical systems, or that
there are fewer GCs per galaxy so the noise reduces the signal.

%%%%%%%%%%%%%%%%%%%%%%%%%%%%%%%%%%%%%%%%%%%%%%%%%%%%%%%%%%%%%%%%%%%%%%
\begin{deluxetable}{c|ccc}
%\centering
%\begin{tabular}
\tablewidth{0pt}
\tablecaption{Kolmogorov-Smirnov probabilities for Binned Galaxies
  Samples.\label{kst}}
\tablehead{
\colhead{$M_z$~range} &
\colhead{$prob_\mathrm{all}$} &
\colhead{$prob_\mathrm{blue}$} &
\colhead{$prob_\mathrm{red}$}
}
\hline
\startdata
$-24,-22$   & 0.004$\pm$0.010& 0.155$\pm$0.122& 0.035$\pm$0.038\\
$-22,-21$   & 0.000$\pm$0.008& 0.096$\pm$0.082& 0.000$\pm$0.022\\
$-21,-20$   & 0.000$\pm$0.000& 0.015$\pm$0.014& 0.000$\pm$0.000\\
$-20,-19$   & 0.377$\pm$0.173& 0.888$\pm$0.244& 0.008$\pm$0.061\\
$-19,-18$   & 0.412$\pm$0.193& 0.303$\pm$0.191& 0.463$\pm$0.234\\
$-18,-17$   & 0.140$\pm$0.115& 0.357$\pm$0.152& 0.085$\pm$0.052\\
$-17,-16$   & 0.120$\pm$0.051& 0.105$\pm$0.047& 0.836$\pm$0.263
\enddata
\tablenotetext{}{Note - $prob_\mathrm{all}$, $prob_\mathrm{blue}$
and $prob_\mathrm{red}$ are the probabilities that the azimuthal 
distribution of GCs is drawn from a random distribution, for all, 
blue and red GCs, respectively. The uncertainty is obtained by
bootstrap. Only galaxies with ellipticities greater than $0.2$ are
included in the bins}
\end{deluxetable}
%%%%%%%%%%%%%%%%%%%%%%%%%%%%%%%%%%%%%%%%%%%%%%%%%%%%%%%%%%%%%%%%%%%%%%

\subsection{Azimuthal Distribution}

In order to further investigate whether the non-random distribution of
GCs has something to do with the host galaxy, we plotted the histogram of GCs as a function of the azimuthal angle from
the galaxy major axis. We
assume that the distribution is axisymmetric in nature, and we define
the major axis of the galaxy to be at
$\phi^\prime = 0^\circ$, and all GCs to have $\phi^\prime$ to be
within the interval $[-90^\circ,90^\circ]$.
Specifically, if a given GC's position angle is out of this interval, we
add or subtract $180^\circ$.
In an axisymetric system, this arrangement doubles the magnitude of
the signal in the peak, if any.

Let us designate the number of observed GCs in a bin by
$N_\mathrm{ob}$ and that of randomized GCs by $N_\mathrm{ran}$ which is the mean in a
bin over the randomized samples. We
define the ``excess'' quantity $E$ as:
\begin{equation}
E=N_\mathrm{ob}-N_\mathrm{ran}
\end{equation}
We assume the number of GCs in a bin observes a
Possion distribution, and we can get the estimate of the deviation
of $E$ where $N$ is the times we do randomization ($N=1000$):

\begin{equation}
\sigma=\sqrt{N_\mathrm{ob}+\frac{N_\mathrm{ran}}{N}}\label{eq2}
\end{equation}

\begin{figure}
\plotone{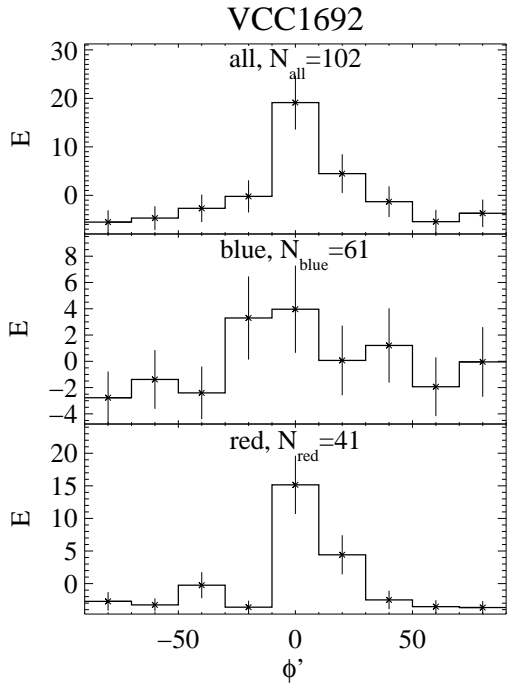}
%\plotone{imgs/mine/e1692}
\caption{VCC1692. The top, middle and bottom plots show the
histograms of all, blue, and red GCs, respectively, versus position
angle (in degrees, east of north from the major axis). The error bar 
shows the uncertainty obtained by the equation (\ref{eq2}). In all
cases, the GC azimuthal distribution shows a strong peak at the
position angle of the major axis, showing that the GCs are aligned
with the light of their host galaxy. \label{e1692}}
\end{figure}

For illustration, we show the azimuthal distribution for VCC~1692 in
Figure~\ref{e1692}.  We find that the red GCs significantly align with the
galaxy's major axis, and the blue GCs also show this trend, but to a
lesser extent.  For most galaxies, the azimuthal histograms are quite
noisy due to the low number of GCs.  Therefore, like in the previous
section, we determine the azimuthal distributions of the binned galaxy
samples.  We select our galaxies and combine the GC systems in the
same way as we did for the K-S test, as shown in Table~\ref{kst}.

\begin{figure*}
\plottwo{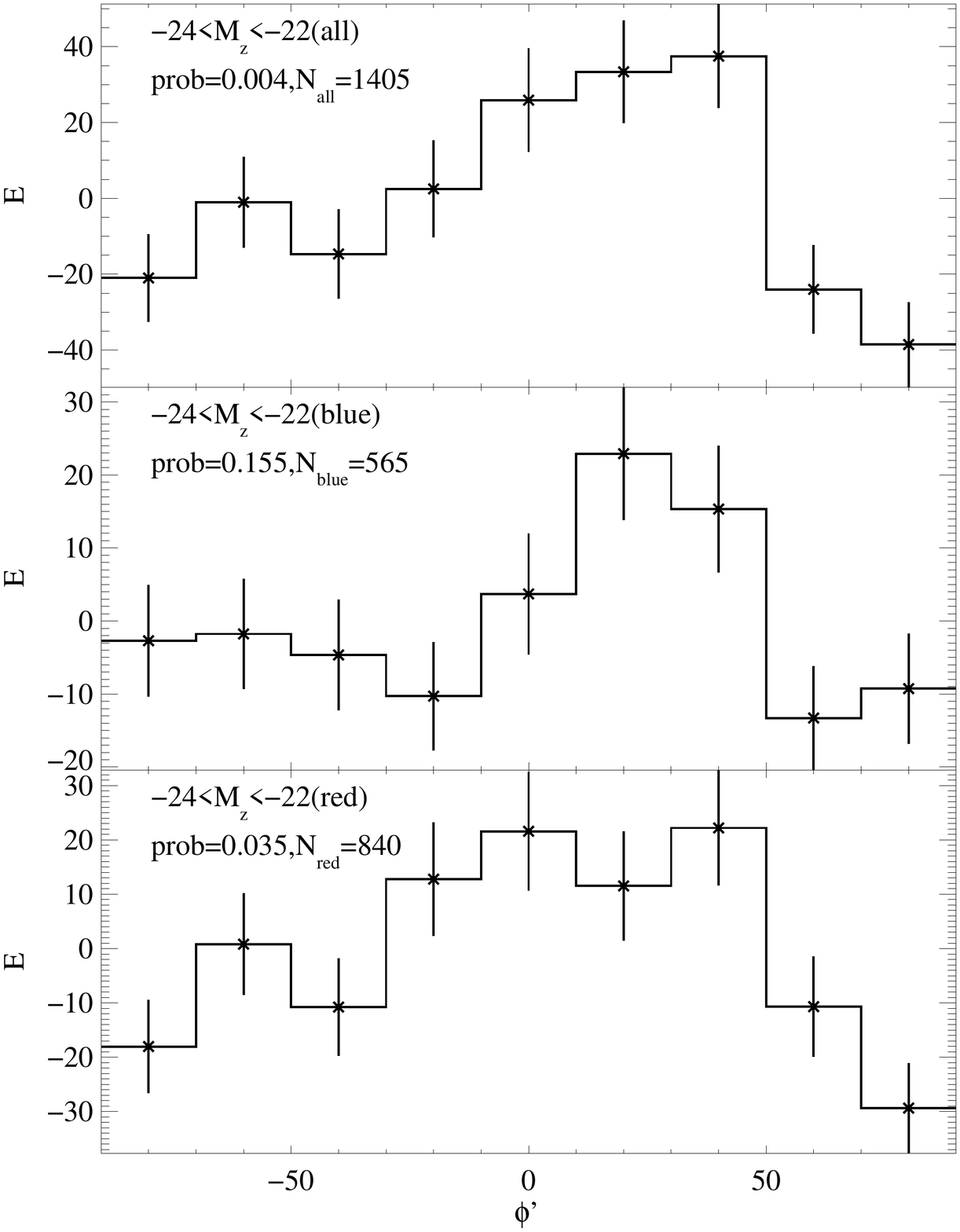}{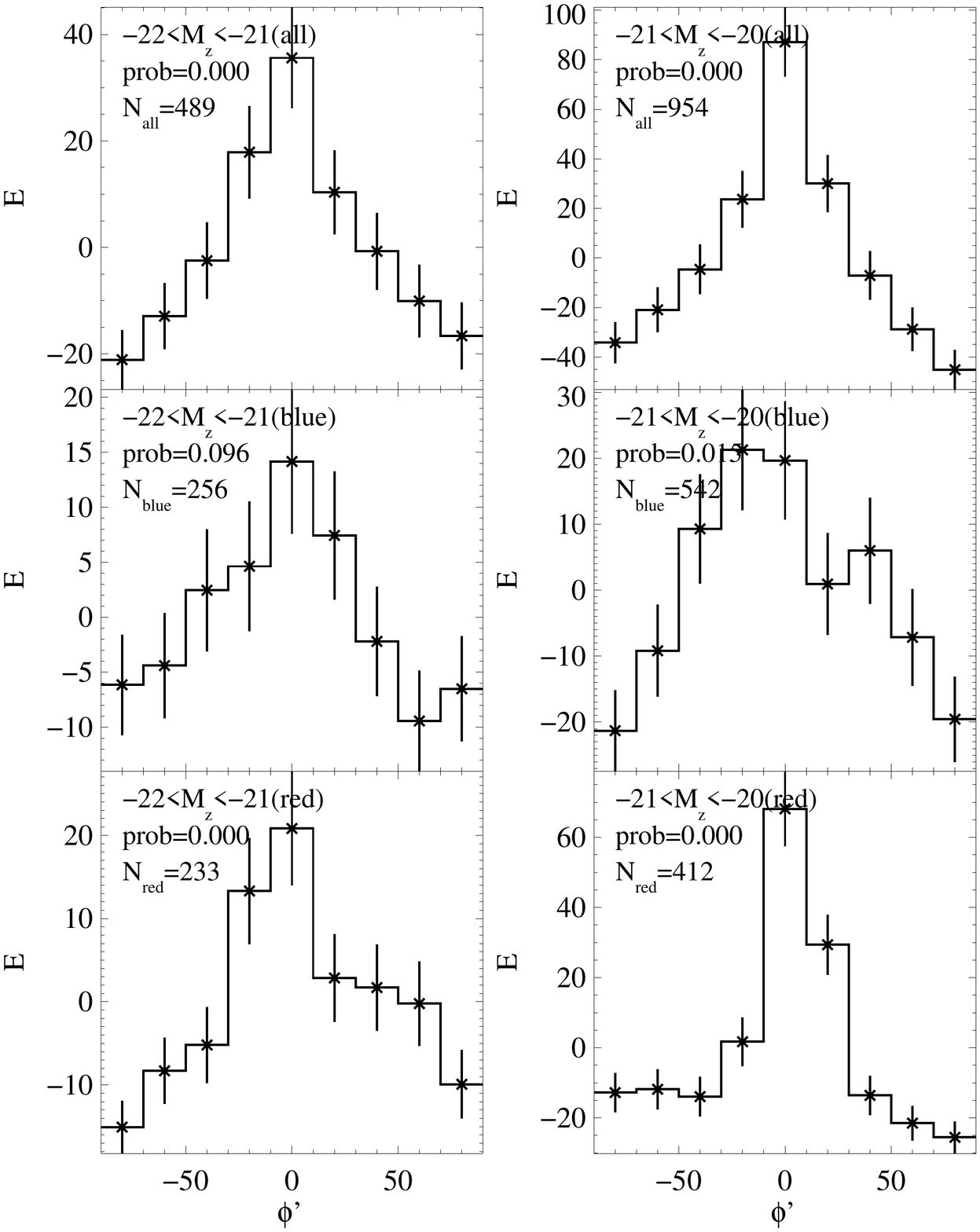}
\caption{Azimuthal distribution of GCs versus position angle (degrees,
  east of north from the major axis) for samples binned by galaxy
  luminosity.  (Left) All galaxies with $\epsilon >0.2$ and $-24<M_z<-22$.
  (Right) Same as left, but for galaxies with $-22<M_z<-21$ and $-21<M_z<-20$.
  The peak at $\phi^\prime=0$~deg means that the GCs are clustered around
  the major axes of their host galaxies. We find this clustering is
  strong for red GCs, but also very significant for blue GCs. For the
  more massive bins, the small field of view may dilute the signal.\label{fig:z7t11}}
\end{figure*}

\begin{figure*}
\plottwo{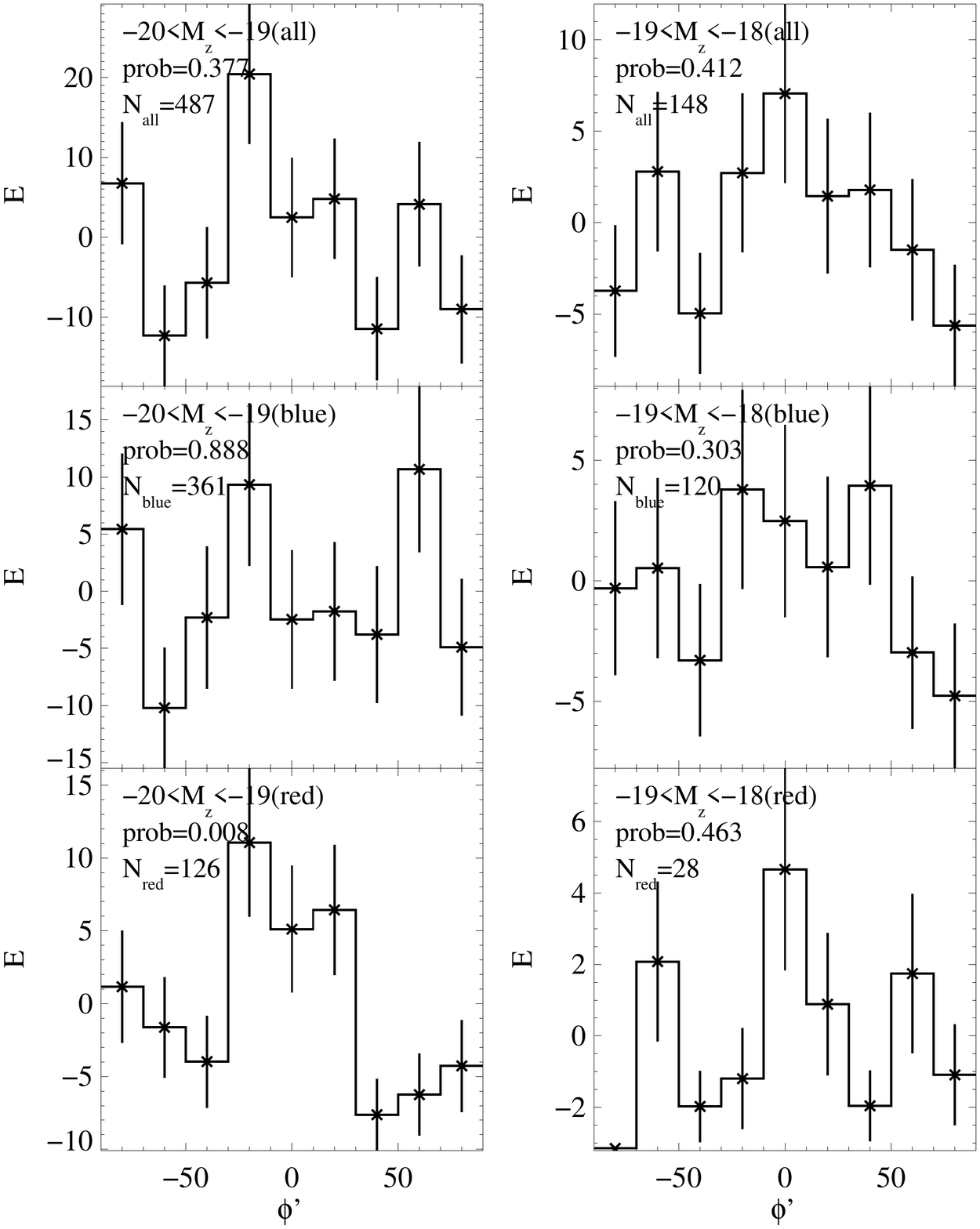}{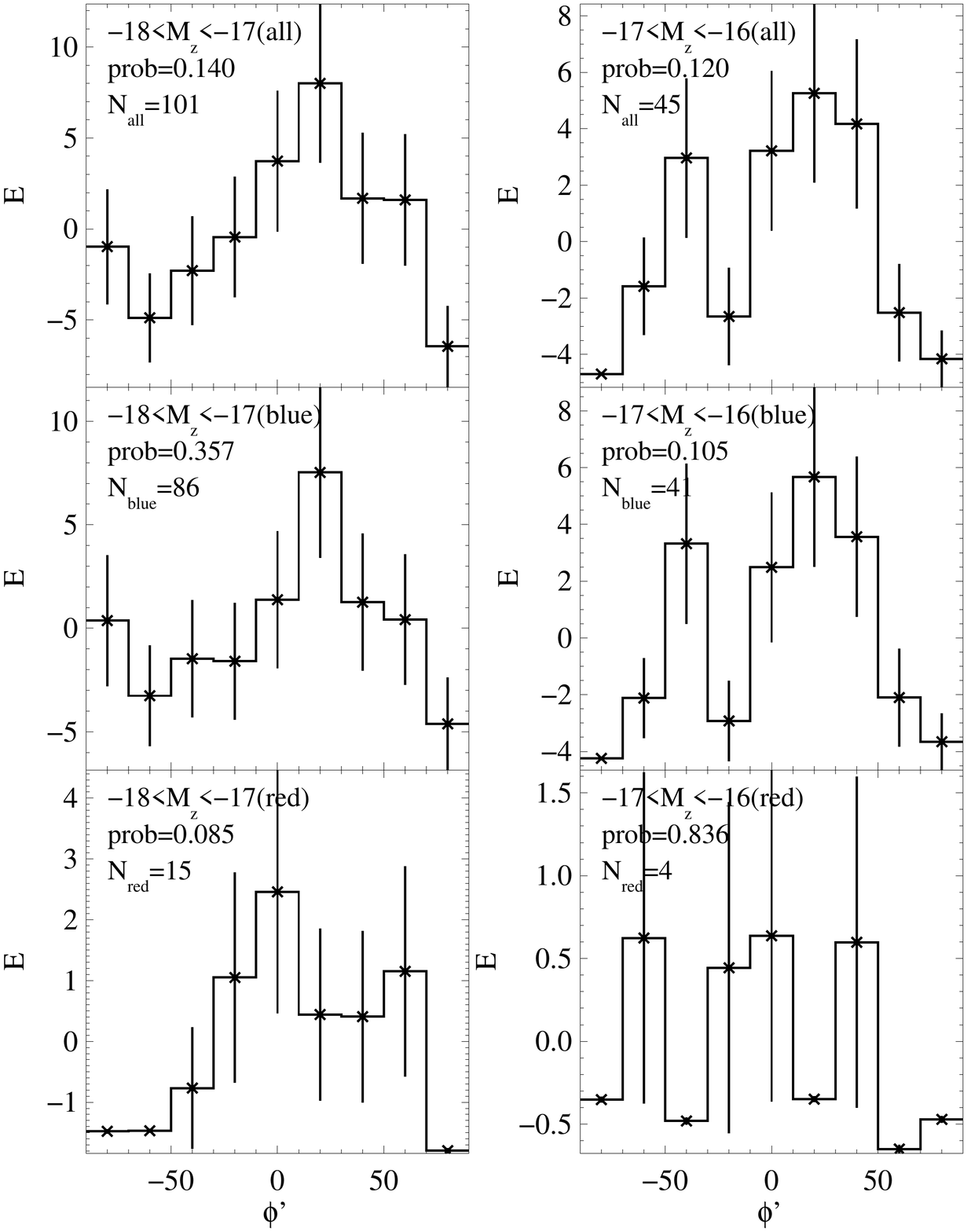}
\caption{Azimuthal distribution of GCs versus position angle (degrees,
  east of north from the major axis) for samples binned by galaxy
  luminosity. All galaxies with $\epsilon >0.2$ and
  $-20<M_z<-19$ and $-19<M_z<-18$ (left), and $-18<M_z<-17$ and 
  $-17<M_z<-16$ (right).  These binned samples show a strong
  clustering around the major axes of galaxies, even for dwarfs. \label{fig:z11t15}}
\end{figure*}

The combined azimuthal GC distributions for the binned samples are plotted
in Figures~\ref{fig:z7t11} and \ref{fig:z11t15}.  As with the K-S
test, we find that the GC azimuthal distribution is significantly
anisotropic for the brighter galaxies.  In this case, we establish that the
GCs are aligned along the major axis of their hosts.  The galaxies
that show this trend most strongly are those that have magnitudes in
the range $-22<M_z<-20$, where it is seen unambiguously for both blue
and red GCs.

%%%%%%%%%%%%%%%%%%%%%%%%%%%%%%%%%%%%%%%%%%%%%%%%%%%%%%%%%%%%%%%%%%%%%%%%%%%%%

%\iffalse
%\begin{figure}[htbp] \centering
%\begin{picture}(100,150)(0,0)
%\put(0,0){\includegraphics[width=4cm,height=4cm]{imgs/mine/zfrom_9_to_11}}
%\end{picture}
%\caption{test\label{zfrom9}}
%\end{figure}
%\fi
%%%%%%%%%%%%%%%%%%%%%%%%%%%%%%%%%%%%%%%%%%%%%%%%%%%%%%%%%%%%%%%%%%%%%%%%%%%%%

%%%%%%%%%%%%%%%%%%%%%%%%%%%%%%%%%%%%%%%%%%%%%%%%%%%%%%%%%%%%%%%%%%%%%%%%%%%%%
%%%%%%%%%%%%%%%%%%%%%%%%%%%%%%%%%%%%%%%%%%%%%%%%%%%%%%%%%%%%%%%%%%%%%%%%%%%%%

%%%%%%%%%%%%%%%%%%%%%%%%%%%%%%%%%%%%%%%%%%%%%%%%%%%%%%%%%%%%%%%%%%%%%%%%%%%%%
%%%%%%%%%%%%%%%%%%%%%%%%%%%%%%%%%%%%%%%%%%%%%%%%%%%%%%%%%%%%%%%%%%%%%%%%%%%%%

\section{Discussion}

Our results indicate that for luminous early-type galaxies with
moderate to high ellipticity, the spatial distribution of GCs is
aligned with the stars in the host galaxies.  For galaxies with small
ellipticity, the GCs tend to be more randomly distributed, but this
neither proves nor disproves the hypothesis that GCs are following the
galaxy light.  The alignment appears strongest in our data for
galaxies around intermediate luminosity ($-22<M_z<-19$).  Early-type
galaxies in this luminosity range are preferentially lenticular or
more elongated ellipticals.  They are also both small enough such that a
significant fraction of their GC system is in the ACS/WFC field of
view, yet luminous enough to still have substantial GC systems.  They
are in many ways the best targets for this study, and we see a clear
signal for both the red and blue GCs.

The red, metal-rich GCs show the strongest alignment with the major axis of
their host galaxies.  These GCs are often hypothesized to be
formed in same starburst events that formed the bulk of the galaxy
(e.g., Ashman \& Zepf 1992), so it is unsurprising that they
should be closely associated with the bulk of the star light.  On the
other hand, it is not necessary for them to have been formed together,
as metal-rich GCs could have been accreted along the same preferred
axis.  For lenticular (S0) galaxies that have both bulges and stellar disks, it
is interesting to ask whether the red GCs seem to be more associated
with bulges or disks.  In some more edge-on galaxies where we can hope
to distinguish disk populations, such as VCC~1692 shown in
Figure~\ref{fig:1692zoom}, the GCs do appear to be more disky in their
distribution.  VCC~1062, classified as an SB0, is another example
where the red GCs almost exactly trace the disk light.  In some
early-type galaxies, the metal-rich GCs appear to be more
associated with the stellar disk than they are with the bulge.
\footnote{Previously, Peng \etal (2006b) found that
diffuse star clusters in the ACSVCS galaxies tend to be found in
lenticular galaxies and were spatially correlated with their disks.
The GCs we study in this paper are different from the diffuse star clusters and have
magnitudes and sizes typical of GCs.}

We also find that the blue, metal-poor GCs in the inner regions of luminous early-type
galaxies also align with the major axes of their hosts.  This result
is more surprising, as the canonical metal-poor halo system is usually
pictured as spherical.  The anisotropy in the distribution of blue GCs is not as
pronounced as that for the red GCs, but it is still detected at high
significance.  This is interesting, as metal-poor GCs often contain
some of the oldest stars in the galaxy, and may form in some of the
earliest collapsing dark matter overdensities (Moore \etal 2006).
Metal-poor halo stars and GCs may also be the best stellar tracer of the
overall matter distribution (e.g., Abadi, Navarro, \& Steinmetz 2006).
That the blue GCs in many of the luminous galaxies appear to have
anisotropic  azimuthal distributions is a clue that these galaxies did
not form their old halos in an environment of isotropic merging, but
perhaps were created along local filaments.  Brainerd (2005) used the
SDSS to find that satellite galaxies within $r_p\lesssim100$~kpc are
also preferentially aligned along the major axis.  If some
blue GCs have their origin in merging satellites, then this is a
suggestive connection.  The connection between satellite dwarfs and halo GCs has been
made in kinematic studies (\cote\ \etal 2001, Woodley 2006), where the
rotation axes of metal-poor GCs and satellite dwarfs were found to be
aligned. Furthermore, a recent GC kinematic study of M87, the second
most luminous galaxy in our sample, indicates that its GC system was
possibly affected by a recent merger (Strader \etal 2011; Romanowsky \etal 2012).

The field of view of our data is too small to evaluate the shape of
the entire GC system for the luminous galaxies, but it would be very
interesting if future wide-field data could extend these studies to
larger radii. If metal-poor GCs trace the total mass distribution,
then this may give us a visible tracer of the flattening of dark
matter halos in these galaxies. Given that these galaxies reside in a dense cluster 
environment, the shapes of their halos and their
GC systems may be unlike that of a galaxy like the Milky Way.
The tendency for the more luminous galaxies to show this alignment is
similar to the trends seen in the alignment between galaxies their
surrounding large scale structure (West \& Blakeslee 2000; Wang \etal 2008; Faltenbacher \etal 2009).

The ACS field of view necessarily limits us to studying the ``inner
halo'' populations of the more luminous galaxies.  Wide-field imaging,
such as that which will be provided by the Next Generation Virgo
Cluster Survey (NGVS; Ferrarese \etal 2012), will provide information
on the possible alignment of GCs in the outer regions of these
galaxies. 

\section{Conclusions}
We have used data from the ACS Virgo Cluster Survey to study the
azimuthal distribution of globular clusters around early-type
galaxies.  We use the Kolmogorov-Smirnov test in conjunction with
tailored control samples to determine the degree of anisotropy in the
azimuthal GC distribtions of the sample galaxies.  We also combine the
GC systems of galaxies with similar luminosities.  We find that:

\begin{itemize}

\item The azimuthal distribution of GCs is strongly anisotropic around
  early-type galaxies with moderate to high ellipticity
  ($\epsilon>0.2$) and intermediate to high luminosity ($M_z<-19$).
  We see no strong trends for dwarf galaxies.

\item In these galaxies, the GCs are preferentially aligned along the
  major axis of the host galaxy.

\item The red GCs exhibit the strongest correlation with galaxy light,
  and in some cases may be associated with the stellar disks of
  lenticulars.  This association strengthens the idea that the
  formation of red GCs is linked to that of the metal-rich field star
  population.

\item The blue GCs in these galaxies also show a significant tendency
  to be aligned with the host galaxy major axis, although to a lesser
  extent than the red GCs.  That the blue GC distributions are also
  non-spherical in their distribution suggests that these galaxies did not
  experience mergers from all directions, and instead formed their
  halos along a direction defined by their current major axis.

\end{itemize}

\acknowledgments
E.~W.~P. gratefully acknowledges support from the Peking
University Hundred Talent Fund (985) and grants 10873001 and 11173003 from the
National Natural Science Foundation of China.  A.~J. acknowledges support from
BASAL CATA PFB-06  and the Millennium Science Initiative, Chilean
Ministry of Economy (Nucleus P07-021-F). The authors thank the
anonymous referee for comments that 
improved the paper. 

This research has made use of the NASA/IPAC Extragalactic Database
(NED) which is operated by the Jet Propulsion Laboratory, California
Institute of Technology, under contract with the National Aeronautics
and Space Administration.

Facilities: \facility{HST(ACS)}

%%%%%%%%%%%%%%%%%%%%%%%%%%%%%%%%%%%%%%%%%%%%%%%%%%%%%%%%%%%%%%%%%%%%%%%%%%%%%
%\begin{figure*}[htbp]
%\plotone{imgs/mine/efrom0}
%
%\caption{efrom0\label{efrom0}}
%\end{figure*}
%%%%%%%%%%%%%%%%%%%%%%%%%%%%%%%%%%%%%%%%%%%%%%%%%%%%%%%%%%%%%%%%%%%%%%%%%%%%%
%
%%%%%%%%%%%%%%%%%%%%%%%%%%%%%%%%%%%%%%%%%%%%%%%%%%%%%%%%%%%%%%%%%%%%%%%%%%%%%%
%\begin{figure*}[htbp]
%\plotone{imgs/mine/efrom8}
%
%\caption{efrom0\label{efrom8}}
%\end{figure*}
%%%%%%%%%%%%%%%%%%%%%%%%%%%%%%%%%%%%%%%%%%%%%%%%%%%%%%%%%%%%%%%%%%%%%%%%%%%%%%

%%%%%%%%%%%%%%%%%%%%%%%%%%%%%%%%%%%%%%%%%%%%%%%%%%%%%%%%%%%%%%%%%%%%%%%%%%%%%%
%\begin{figure*}[htbp]
%\plotone{imgs/mine/efrom20}
%
%\caption{efrom0\label{efrom20}}
%\end{figure*}
%%%%%%%%%%%%%%%%%%%%%%%%%%%%%%%%%%%%%%%%%%%%%%%%%%%%%%%%%%%%%%%%%%%%%%%%%%%%%%

%%%%%%%%%%%%%%%%%%%%%%%%%%%%%%%%%%%%%%%%%%%%%%%%%%%%%%%%%%%%%%%%%%%%%%%%%%%%%
%\begin{figure*}[htbp]
%\plotone{imgs/mine/efrom32}
%
%\caption{efrom0\label{efrom32}}
%\end{figure*}
%%%%%%%%%%%%%%%%%%%%%%%%%%%%%%%%%%%%%%%%%%%%%%%%%%%%%%%%%%%%%%%%%%%%%%%%%%%%%

%\begin{figure*}[htbp]
%\plotone{imgs/mine/efrom44}
%
%\caption{efrom0\label{efrom44}}
%\end{figure*}
%\begin{figure*}[htbp]
%\plotone{imgs/mine/efrom56}
%
%\caption{efrom0\label{efrom56}}
%\end{figure*}
%\begin{figure*}[htbp]
%\plotone{imgs/mine/efrom68}
%
%\caption{efrom0\label{efrom68}}
%\end{figure*}
%\begin{figure*}[htbp]
%\plotone{imgs/mine/efrom80}
%
%\caption{efrom0\label{efrom80}}
%\end{figure*}

%\clearpage

{}

\clearpage

\clearpage

%% Tables may also be prepared as separate files. See the accompanying
%% sample file table.tex for an example of an external table file.
%% To include an external file in your main document, use the \input
%% command. Uncomment the line below to include table.tex in this
%% sample file. (Note that you will need to comment out the \documentclass,
%% \begin{document}, and \end{document} commands from table.tex if you want
%% to include it in this document.)

%\clearpage

%% The following command ends your manuscript. LaTeX will ignore any text
%% that appears after it.

\end{document}